\documentclass[conference]{IEEEtran}

\usepackage[T1]{fontenc}
\usepackage{inconsolata}

\usepackage{alltt}
\usepackage{graphicx}
\usepackage{xspace}
\usepackage{subfigure}
\usepackage{algorithm}
\usepackage{algpseudocode}
\usepackage{amssymb}
\usepackage{xcolor}
\usepackage{listings}
\usepackage{hyperref}
\usepackage{framed}
\usepackage{MnSymbol}

\newif\ifsubmit

\submittrue

\definecolor{mygray}{gray}{0.90}
\definecolor{mydarkgray}{gray}{0.30}
\definecolor{mygreen}{rgb}{0.2,0.7,0}
\definecolor{mymauve}{rgb}{0.58,0,0.82}
\definecolor{myblue}{rgb}{0,0,0.7}
\definecolor{myred}{rgb}{0.7,0.2,0}

\makeatletter
\lst@AddToHook{TextStyle}{\let\lst@basicstyle\ttfamily\small\fontfamily{pcr}\selectfont}
\makeatother

\newcommand{\lst}[1]{\lstinline{#1}}
\newcommand{\dash}{DASH\xspace}
\newcommand{\dart}{DART\xspace}

\newcommand{\stdvector}{\lstinline{std::vector}\xspace}
\newcommand{\stddeque}{\lstinline{std::deque}\xspace}
\newcommand{\stdsort}{\lstinline{std::sort()}\xspace}
\newcommand{\stdfill}{\lstinline{std::fill()}\xspace}

\newcommand{\dasharray}{\lstinline{dash::Array}\xspace}
\newcommand{\dashmatrix}{\lstinline{dash::Matrix}\xspace}
\newcommand{\dashnarray}{\lstinline{dash::NArray}\xspace}
\newcommand{\dashteamall}{\lstinline{dash::Team::All()}\xspace}
\newcommand{\globref}[1]{\lstinline{GlobRef<#1>}\xspace}
\newcommand{\globptr}[1]{\lstinline{GlobPtr<#1>}\xspace}
\newcommand{\globiter}[1]{\lstinline{GlobIter<#1>}\xspace}
\newcommand{\globmem}[1]{\lstinline{GlobMem<#1>}\xspace}
\newcommand{\pattern}[0]{\lstinline{Pattern}\xspace}
\newcommand{\dashpattern}[0]{\lstinline{dash::Pattern}\xspace}
\newcommand{\team}[0]{\lstinline{Team}\xspace}
\newcommand{\teamsplit}[1]{\lstinline{team.split(#1)}\xspace}
\newcommand{\dashteam}[0]{\lstinline{dash::Team}\xspace}
\newcommand{\dashminelement}[1]{\lstinline{dash::min\_element(#1)}\xspace}

\newcommand{\init}[0]{\lstinline{init()}\xspace}
\newcommand{\finalize}[0]{\lstinline{finalize()}\xspace}
\newcommand{\myid}[0]{\lstinline{myid()}\xspace}
\newcommand{\size}[0]{\lstinline{size()}\xspace}

\newcommand{\dashteamallmyid}{\lstinline{dash::Team::All().myid()}\xspace}
\newcommand{\dashteamallbarrier}{\lstinline{dash::Team::All().barrier()}\xspace}

\newcommand{\dashbarrier}{\lstinline{dash::barrier()}\xspace}

\newcommand{\none}[0]{\lstinline{NONE}\xspace}
\newcommand{\cyclic}[0]{\lstinline{CYCLIC}\xspace}
\newcommand{\blocked}[0]{\lstinline{BLOCKED}\xspace}
\newcommand{\blockcyclic}[1]{\lstinline{BLOCKCYCLIC(#1)}\xspace}

\newcommand{\colmajor}[0]{\lstinline{COL\_MAJOR}\xspace}
\newcommand{\rowmajor}[0]{\lstinline{ROW\_MAJOR}\xspace}

\newcommand{\mdim}[0]{multidimensional\xspace}
\newcommand{\Mdim}[0]{Multidimensional\xspace}
\newcommand{\onedim}[0]{one-dimensional\xspace}
\newcommand{\twodim}[0]{two-dimensional\xspace}

\ifsubmit

\newenvironment{done}{}{}

\else

\colorlet{shadecolor}{green!20}
\newenvironment{done}{\begin{snugshade}}{%
    \end{snugshade}\ignorespacesafterend%
}

\fi

\begin{document}

\title{DASH: A C++ PGAS Library for Distributed Data Structures and Parallel Algorithms}

\author{
  \IEEEauthorblockN{Karl Fuerlinger, Tobias Fuchs, and Roger Kowalewski}

  \IEEEauthorblockA{Ludwig-Maximilians-Universit{\"a}t (LMU)
    Munich,\\ Computer Science Department, MNM
    Team,\\ Oettingenstr.~67, 80538 Munich, Germany\\ Email:
    first.last@nm.ifi.lmu.de} }
\maketitle

\begin{abstract}
We present DASH, a C++ template library that offers distributed data structures and parallel algorithms and implements a compiler-free PGAS (partitioned global address space) approach. DASH offers many productivity and performance features such as global-view data structures, efficient support for the owner-computes model, flexible \mdim data distribution schemes and inter-operability with STL (standard template library) algorithms. DASH also features a flexible representation of the parallel target machine and allows the exploitation of several hierarchically organized levels of locality through a concept of Teams. We evaluate DASH on a number of benchmark applications and we port a scientific proxy application using the MPI two-sided model to DASH. We find that DASH offers excellent productivity and performance and demonstrate scalability up to 9800 cores.
\end{abstract}


\section{Introduction}\label{sec:Introduction}


\begin{done}
The PGAS (Partitioned Global Address Space) model is a promising approach for programming large-scale systems~\cite{Amarasinghe:2009:Exascale, Gomez-Iglesias:2015:Libraries, Yelick:2007:Productivity}. When dealing with unpredictable and irregular communication patterns, such as those arising from graph analytics and data-intensive applications, the PGAS approach is often better suited and more convenient than two-sided message passing~\cite{Jose:2013:Graph500}. The PGAS model can be seen as an extension of threading-based shared memory programming to distributed memory systems, most often employing one-sided communication primitives based on RDMA (remote direct memory access) mechanisms~\cite{Almasi:2011:PGAS}. Since one-sided communication decouples data movement from process synchronization, PGAS models are also potentially more efficient than classical two-sided message passing approaches~\cite{Belli:2015:Notified}.

However, PGAS approaches have so far found only limited acceptance and adoption in the HPC community~\cite{Wong:2011:PraceSurvey}. One reason for this lack of widespread usage is that for PGAS \emph{languages}, such as UPC~\cite{UPC:2005:Spec}, Titanium~\cite{Aiken:1998:Titanium}, and Chapel~\cite{Chamberlain:2007:PPC}, adopters are usually required to port the whole application to a new language ecosystem and are then at the mercy of the compiler developers for continued development and support. Developing and maintaining production-quality compilers is challenging and expensive and few organizations can afford such a long-term project.

\emph{Library-based} approaches are therefore an increasingly attractive low-risk alternative and in fact some programming abstractions may be better represented through a library mechanism than a language construct (the data distribution patterns described in Sect~\ref{sect:MultiDimData} are an example). Global Arrays~\cite{Nieplocha:1994:GA} and OpenSHMEM~\cite{Chapman:2010:Introducing} are two popular examples for compiled PGAS libraries with a C API, which offer an easy integration into existing code bases. However, precompiled libraries and static APIs severely limit the productivity and expressiveness of programming systems and optimizations are typically restricted to local inlining of routines.

C++, on the other hand, has powerful abstraction mechanisms that allow for generic, expressive, and highly optimized libraries~\cite{Lee:2002:Generic}. With a set of long awaited improvements incorporated in C++11~\cite{C++:2011:Standard}, the language has recently been used to implement several new parallel programming systems in projects such as UPC++~\cite{Zheng:2014:UPC++}, Kokkos~\cite{Edwards:2014:Kokkos}, and RAJA~\cite{Hornung:2014:RAJA}.

In this paper we describe \dash, our own C++ template library that implements the PGAS model and provides generic distributed data structures and parallel algorithms. DASH realizes the PGAS model purely as a C++ template library and does not require a custom (pre-)compiler infrastructure, an approach sometimes called compiler-free PGAS. Among the distinctive features of \dash are its inter-operability with existing (MPI) applications, which allows the porting of individual data structures to the PGAS model, and support for hierarchical locality beyond the usual two-level distinction between local and remote data.

\end{done}

The rest of this paper is organized as follows. In Sect.~\ref{sec:Overview} we provide a high-level overview of DASH, followed by a more detailed discussion of the library's abstractions, data structures and algorithms in Sect.~\ref{sec:DSandAlgo}. In Sect.~\ref{sec:Evaluation} we evaluate DASH on a number of benchmarks and a scientific proxy application written in C++. In Sect.~\ref{sec:RelatedWork} we discuss related work and we conclude and describe areas for future work in Sect.~\ref{sec:Conclusion} 

\section{An Overview of DASH}\label{sec:Overview}

This section provides a high level overview of DASH and its implementation based on the runtime system DART\@. 

\subsection{DASH and DART}\label{sect:DashAndDart}

\begin{done}

\dash is a C++ template library that is built on top of \dart (the DAsh RunTime), a lightweight PGAS runtime system implemented in C. The \dart interface specifies basic mechanisms for global memory allocation and addressing using global pointers, as well as a set of one-sided put and get operations. The \dart interface is designed to abstract from a variety of one-sided communication substrates such as GASPI~\cite{Grunewald:2013:Gaspi}, GASNet~\cite{Bonachea:2002:Gasnet} and ARMCI~\cite{Nieplocha:1999:ARMCI}. \dash ships with an MPI-3 RMA (remote memory access) based implementation of \dart called DART-MPI~\cite{Zhou:2014:DART} that uses shared memory communicators to optimize intra-node data transfer~\cite{Zhou:2015:SharedMemory}. A single-node implementation utilizing System-V shared memory has also been developed as proof-of-concept, and experimental support for GPUs was added in DART-CUDA~\cite{Zhou:2015:DARTCUDA}.

\end{done}

\subsection{Execution Model}\label{sect:ExecutionModel}

\begin{figure}\centering
\begin{minipage}[t]{0.49\textwidth}
\begin{lstlisting}[frame=lines]
#include <libdash.h>
#include <iostream>
using namespace std;

int main(int argc, char *argv[])
{
  dash::init(&argc, &argv);
  // private scalar and array
  int p; double s[20];|\label{line:private}|
  // globally shared array of 1000 integers
  dash::Array<int> a(1000);|\label{line:shared}|
  // initialize array to 0 in parallel
  dash::fill(a.begin(), a.end(), 0);|\label{line:algo}|
  // global reference to last element
  dash::GlobRef<int> gref = a[999];
  if (dash::myid() == 0) {
    // global pointer to last element
    dash::GlobPtr<int> gptr = a.end() - 1;
    (*gptr) = 42;|\label{line:gptr}|
  }
  dash::barrier();
  cout << dash::myid() << " " << gref << endl;|\label{line:gref}|
  cout << dash::myid() << " " << a[0] << endl;|\label{line:aref}|
  dash::finalize();
}
\end{lstlisting}
\end{minipage}
\caption{A simple stand-alone DASH program illustrating global data structures, global references, and global pointers.}\label{fig:Example}
\end{figure}

\begin{done}
  \dash follows the SPMD (single program, multiple data) model with hierarchical additions. In order to liberate terminology from a concrete implementation choice, we refer to the individual participants in a \dash program as {\bf units} (cf.\ threads in UPC and images in Co-Array Fortran). Units may be implemented as full operating system processes or realized as lightweight threads. The total number of units is determined when the program is launched and stays unchanged throughout the program's lifetime. Units are organized into {\bf teams} that can be dynamically created and destroyed at runtime. The sole method to establish a new team is to create a subset of an existing team starting from \dashteamall, the built-in team representing all units in a program.

In the \dash programming model, teams form the basis for all collective synchronization, communication, and memory allocation operations. Constructors for \dash data structures have an optional \team parameter that defaults to \dashteamall. Since a team represents physical resources (the set of memory allocations performed by the team members and their execution capabilities), \dashteam is implemented as a move-only type that cannot be copied.
\end{done}

Fig.~\ref{fig:Example} shows a simple stand-alone DASH program. \init initializes the runtime, \finalize reclaims resources. \myid is shorthand for \dashteamallmyid and returns the zero-based ID of the calling unit. Similarly, \size returns the number of participating units. \dashbarrier is shorthand for \dashteamallbarrier and synchronizes all units.

\subsection{Memory Model}\label{sect:MemoryModel}

\begin{done}
\dash implements a \emph{private-by-default} memory model where regular C++ variables and STL (standard template library) containers are private and cannot be accessed by other units (Fig.~\ref{fig:Example}, line~\ref{line:private}). To make data \emph{shared} and accessible by other units, the containers provided by \dash (such as \dasharray and \dashmatrix) are allocated over a team (line~\ref{line:shared}). The team members provide the memory and can later access the shared data via one-sided put and get operations that are typically triggered automatically in response to higher-level access operations (see Sect.~\ref{sect:Referencing}).

The sources and targets for one-sided put and get operations are specified in terms of global pointers. A DART global pointer identifies a location in global shared memory and consists of a unit ID, a memory segment identifier, and an offset within this segment. DART global pointers are 16 bytes in size and can address any 64 bit memory location on up to $2^{32}$ units. The remaining bits are used for flags or reserved for future use.
\end{done}

\subsection{Referencing and Accessing Data}\label{sect:Referencing}

\begin{done}
When using an STL container, such as \stdvector, accessor
functions (\lst{.at()} and the subscript operator\lst{[]}) return
a reference to the stored value. C++ references can be
thought of as named aliases, implemented using memory addresses (i.e., pointers),
with additional safety guarantees enforced by the compiler.
Since a \dash container holds elements that
may not reside in a unit's local memory, data access cannot happen
by C++ reference. Instead, accessor functions return a
{\bf global reference} proxy object of type \globref{T} where \lst{T}
is the element's data type.
\end{done}

\begin{figure}\centering
\includegraphics[width=0.4\textwidth]{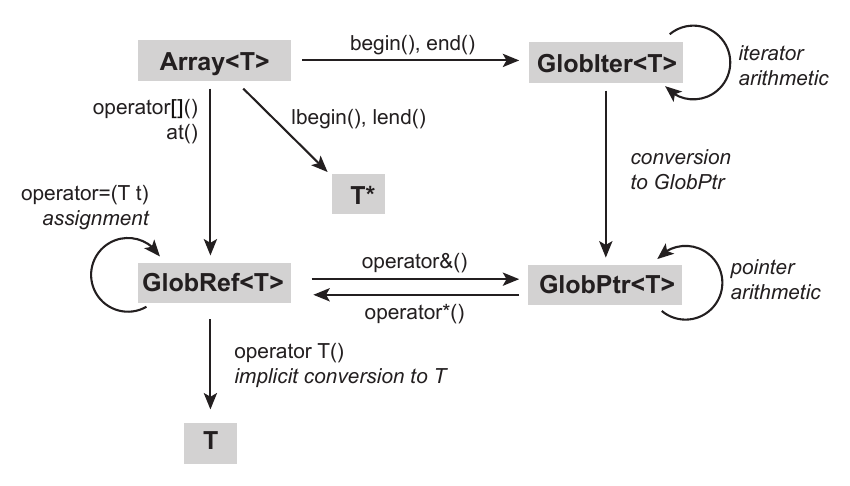}
\caption{The interplay of central abstractions in DASH.}\label{fig:Internals}
\end{figure}

\begin{done}
\globref{} mimics C++ references and behaves in the following way: A \globref{T} is implicitly convertible to a variable of type \lst{T}. For example, given the definition of \lst{gref} in Fig.~\ref{fig:Example},
\lst{cout<<gref} in line~\ref{line:gref} will output 42. Whenever a conversion of global reference to value type is requested, and the global reference denotes a remote location, a \lst{get} operation is performed and the remote value is fetched. If the location is local, the value is directly accessed in shared memory. Conversely, \globref{} implements
an assignment operator for objects of type \lst{T} that performs a \lst{put} operation
of the supplied value to the global memory location referenced by
\globref{}.

A {\bf global pointer} object \globptr{T} is a thin wrapper
around the global pointer provided by DART\@. The main function of the
global pointer is to specify the global memory locations for
one-sided put and get operations. Dereferencing a
global pointer (line~\ref{line:gptr} in Fig.~\ref{fig:Example}) creates
a \globref{T} object, thus \lst{(*gptr)=42} sets the value of the
last array element to 42.
\globptr{} also supports pointer arithmetic and subscripting,
but this only acts on the address part of the global pointer,
while the unit ID remains unchanged. In other words, a global pointer does
not have any \emph{phase} information associated with it and cannot be used to iterate
over a distributed array directly. {\bf Global iterators} are used for that purpose instead.
A global pointer \globptr{T} can be converted to a regular pointer (\lst{T*}) if
the memory is local, otherwise \lst{nullptr} is returned.

A \globiter{T} behaves like a random access iterator that can
iterate over a chunk of global memory by keeping an internal
integer index that can be dynamically converted on-demand to a \globptr{T}.
To realize this index-to-\globptr{} conversion, the \globiter{T} constructor
takes two arguments: A \globmem{T} object that represents a chunk of global memory
and a \pattern object. A \pattern is the \dash approach to express data distribution and memory layout, more details about \mdim patterns is provided in Sect.~\ref{sect:MultiDimData}.

A schematic illustration of the interplay of global pointers, iterators, and references in shown in Fig.~\ref{fig:Internals}. Note, however, that DASH users don't usually need to know or care about these implementation details. Instead, DASH is used with an interface that is familiar to most C++ developers: containers that offer subscripting and iterator-based access, and algorithms working on ranges delimited by iterators (Fig.~\ref{fig:Example}, lines~\ref{line:algo} and~\ref{line:aref}).

\end{done}

\subsection{Teams}\label{sect:Teams}

A \team in DASH is simply an ordered set of units. A new team is always formed as a subset of an existing team, and thus a hierarchy from leaf team to the root (\dashteamall) is maintained. The simplest operation to create a new team is \teamsplit{n}, which creates $n$ new teams, each with an approximately equal number of units. Teams are used to represent hierarchical configurations that arise in the hardware topology or may come from the algorithmic design of the application~\cite{Kamil:2013:Hierarchical}. To reflect the machine topology, an equal-sized split will generally be inadequate. DASH thus computes so-called Locality Domain Hierarchies by integrating information from a number of sources such as PAPI, hwloc~\cite{broquedis2010hwloc} and the OS. Using this mechanism it is, for example, possible to split \dashteamall into sub-teams representing the shared memory nodes on the first level and to then perform another split into sub-teams corresponding to NUMA domains on the second level. This scheme also supports hardware accelerators such as GPUs or Xeon Phi cards and DASH allows the formation of a sub-team that consists of all Xeon-Phi co-processors allocated to an application run.

These hardware-aware teams can then be used for (static) load balancing by identifying the hardware capabilities of each sub-team and adjusting the number of elements accordingly. 


\section{Data Structures and Algorithms in DASH}\label{sec:DSandAlgo}

In this section we describe the fundamental data container offered by dash, the \dasharray. We discuss the flexible data distribution schemes in one and multiple dimensions and the algorithms offered by DASH. 

\subsection{The DASH Array}\label{sect:DASHArray}

\begin{figure}
\centering
\includegraphics[width=0.30\textwidth]{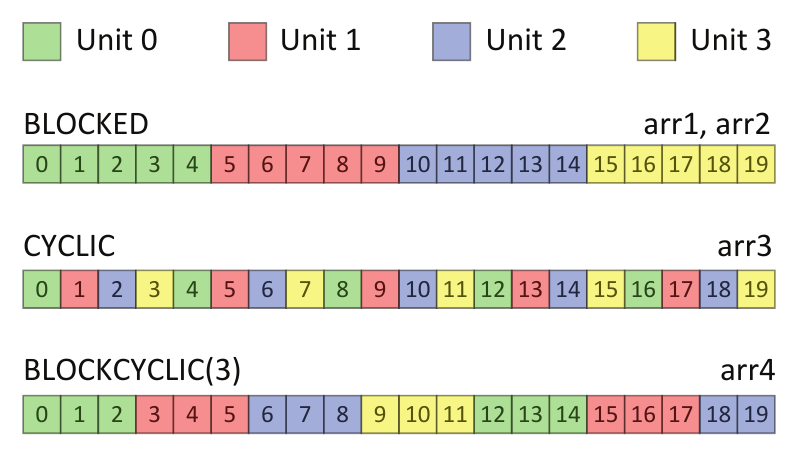}
\caption{The \onedim data distribution patterns supported by DASH.}\label{fig:Pattern1D}
\end{figure}

\begin{done}

The DASH array (\dasharray) is a generic, fixed-size, \onedim container class template, similar in functionality to the built-in arrays that most programming languages offer (for process-local data) and the UPC \emph{shared array} (for distributed data). Once constructed, the size of the array is fixed and cannot be changed. A \dasharray is always constructed over a team and all units in the team contribute an equal amount of memory to hold the array's data. The team used for the allocation is specified as an optional constructor parameter. If no team is explicitly given, the default team, \dashteamall is used. For example:
\begin{lstlisting}[numbers=none,xleftmargin=0cm]
// globally shared array of 1000 integers
dash::Array<int> arr1(1000);

dash::Team& t1 = ...; // construct a new team
// arr2 is allocated over team t1
dash::Array<int> arr2(1000, t1);
\end{lstlisting}

The construction of a DASH array is a collective operation. All units have to call the constructor with the same arguments and it is an error if different arguments are supplied. Besides the template parameter that specifies the type of the container elements, the array constructor takes at least one argument: the total number of elements in the array (its global size). The default data distribution scheme used by \dash is \mbox{\blocked,} which means that each unit stores at most one contiguous block of elements of size $N_{elements}/N_{units}$ rounded up to the next integer.

Optionally, one of the distribution specifiers \blocked, \mbox{\cyclic{},} \blockcyclic{} can be supplied explicitly, where \cyclic is an alias for \blockcyclic{1}. As an example, when run with four units, the following declarations give rise to the distribution patterns shown in Fig.~\ref{fig:Pattern1D}.

\begin{lstlisting}[numbers=none,xleftmargin=0cm]
dash::Array<int> arr1(20); // default: BLOCKED

dash::Array<int> arr2(20, dash::BLOCKED)
dash::Array<int> arr3(20, dash::CYCLIC)
dash::Array<int> arr4(20, dash::BLOCKCYCLIC(3))
\end{lstlisting}
\end{done}

\begin{done}
{\bf Accessing the elements of a \dasharray}. There are various ways in which elements in a \dash array can be accessed. \dash implements a \emph{global-view} PGAS approach in which global data structures are accessed by global indices and iterators. In other words, the expression \lst{a[77]} refers to the same element in the array \lst{a}, regardless of which unit evaluates the expression. Global-view programming has the appealing property that standard sequential algorithms can be used directly with the dash array. For example, \lst{std::sort(a.begin(),a.end())} will employ a standard sequential sort algorithm to sort the elements in the dash array.

Global element access is supported by accessor functions (\lst{at()} and \lst{operator[]()}) and through global iterators. Following established STL conventions, \lst{arr.begin()} returns a global iterator to the first element and \lst{arr.end()} is an iterator to one-past-the-last element in the array \lst{arr}. Thus, \dasharray works seamlessly with the C++11 range-based for loops, so \lst{for(auto el: arr) cout << el;} prints all elements of \lst{arr}.

For performance reasons it is critically important to take locality into account when working with data and all PGAS approaches support an explicit notion of data locality in some form. DASH adopts the concept of {\bf local view proxy objects} to express data locality on the unit level. In addition to this standard two-level differentiation (local vs.\ remote) \dash also support a more general hierarchical locality approach. This is discussed in more detail in Sect.~\ref{sect:Teams}.

The local proxy object \lst{arr.local} represents the part of the array \lst{arr} that is stored locally on a unit (i.e., the local view of the array). \lst{arr.local.begin()} and \lst{arr.local.end()}, or alternatively \lst{arr.lbegin()}, and \lst{arr.lend()} provide raw pointers to the underlying storage for maximum performance. These raw pointers can be used to interface with existing software package such as mathematical libraries. Note that the local proxy object does not respect the global element ordering (as specified by an array's pattern). \lst{local[2]} is simply the third element stored locally and depending on the pattern it will correspond to a different global element. If this global information is required, a pointer to the global proxy object can be converted to a global pointer and to a global iterator.
\end{done}

\begin{figure}
\centering
\includegraphics[width=0.30\textwidth]{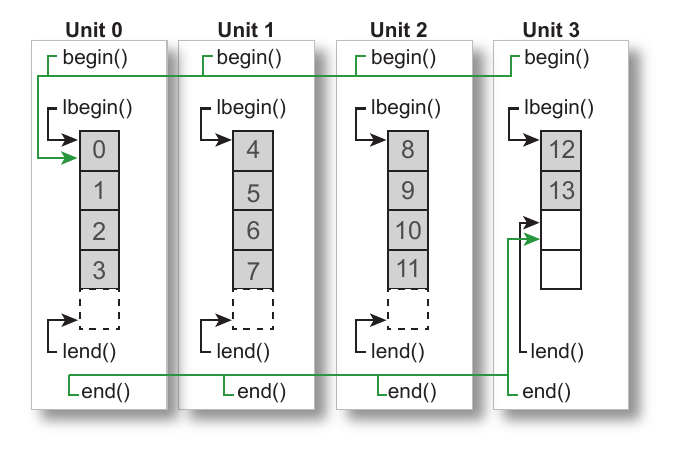}
\caption{Distributed memory layout for a DASH Array with 14 elements distributed over four units. The array supports global-view iteration using begin() and end() as well as local view iteration using lbegin() and lend().}\label{fig:Array}
\end{figure}

\begin{done}
Fig.~\ref{fig:Array} illustrates this concept for a distributed array with 14 elements distributed over four units. Each unit gets a block of four elements, the last unit's storage is underfilled with only two elements. \lst{begin()} and \lst{end()} return the same global iterator for each unit, \lst{lbegin()} and \lst{lend()} return unit-local begin and end iterators, respectively.
\end{done}

\subsection{Working with \Mdim Data}\label{sect:MultiDimData}

\begin{figure}
\centering
\includegraphics[width=0.40\textwidth]{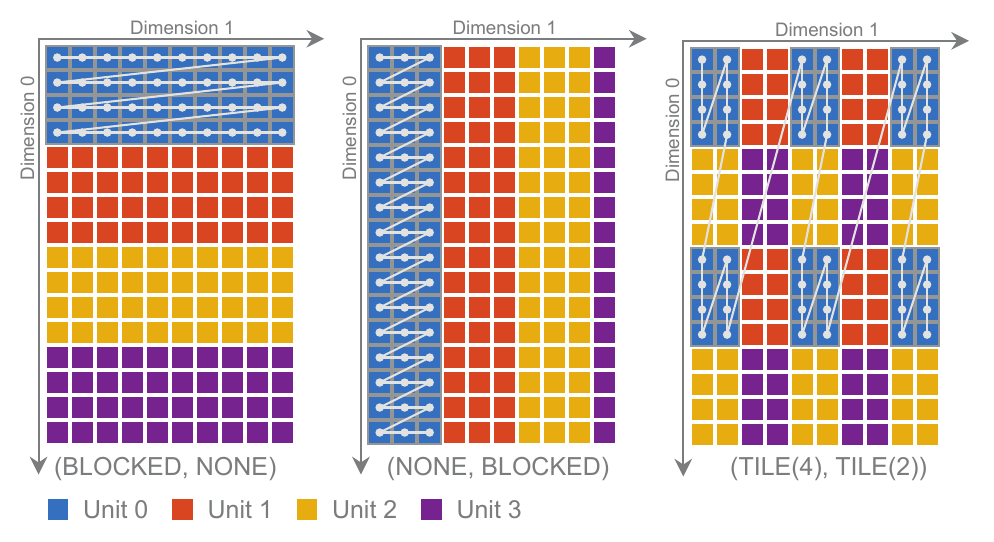}
\caption{Visualization of three $16\times10$ 2D data distribution patterns available in DASH using four units. The different colors correspond to the different units. For unit 0 the visualization additionally included the memory storage order (white dots and connecting lines).}\label{fig:2DPattern}
\end{figure}

\begin{done}
Besides the \onedim array described in Sect.~\ref{sect:DASHArray}, DASH also supports efficient and productive work with \mdim data by providing the DASH N-dimensional array (available as \dashnarray and the alias \mbox{\dashmatrix).} \dashmatrix is a distributed N-dimensional array container class template. Its constructor requires at least two template arguments, one for the element type (\lstinline{int}, \lstinline{double}, etc.) and one for the dimension (N). The following example creates a \twodim integer matrix with 40 rows and 30 columns distributed over all units.
\begin{lstlisting}[numbers=none,xleftmargin=0cm]
dash::Matrix<int, 2> matrix(40, 30); // 1200 elements
\end{lstlisting}

Just like the distributed 1D array, \dashmatrix offers efficient global and local access methods (using coordinates, linear indices, and iterators) and allows for slicing and efficient construction of block regions and lower-dimensional sub-matrix views. Details about the DASH \mdim array and a case study implementing linear algebra routines with performance results rivaling highly tuned linear algebra packages are presented in a publication under review~\cite{Fuchs:2016:Matrix}.

As an extension to the \onedim case, DASH allows the specification of \mdim data distribution patterns. The distribution specifiers \cyclic, \blockcyclic{}, \blocked, and \none can be used in one more dimensions and \none means that no distribution is requested in a particular dimension.  The following example creates two 2D patterns, each with $16\times10$ elements. The resulting pattern (assuming four units) is visualized in Fig.~\ref{fig:2DPattern} (left and middle).
\begin{lstlisting}[numbers=none,xleftmargin=0cm]
dash::Pattern<2> pat1(16, 10, BLOCKED, NONE);
dash::Pattern<2> pat2(16, 10, NONE, BLOCKED);
\end{lstlisting}

\end{done}

\begin{done}
In addition, DASH supports \emph{tiled} patterns where elements are distributed in contiguous blocks of iteration order. Additionally for any \mdim pattern the memory layout (or storage order) can be specified using the \rowmajor or \colmajor template arguments. If not explicitly specified, the default is row major storage. Fig~\ref{fig:2DPattern} (right) shows the following tiled pattern with column major storage order as an example:
\begin{lstlisting}[numbers=none,xleftmargin=0cm]
// arrange team in a 2x2 configuration
dash::TeamSpec<2> ts(2,2);

// 4x2 element tiles, column major layout
dash::TilePattern<2, COL_MAJOR> |$\rhookswarrow$|
      pat3(16, 10, TILE(4), TILE(2), ts);
\end{lstlisting}

Note that \dashnarray and \dashpattern support arbitrarily large dimensions (barring compiler limitations when instantiating the templates) but provide specializations and optimizations for the common two- and three-dimensional cases.
\end{done}

\subsection{DASH Algorithms}

\begin{done}
DASH is first and foremost about data structures, but data structures by themselves are of limited use. Of course, developers are interested in running some computation on the data to achieve results. Often this computation can be composed of smaller algorithmic building blocks. This concept is supported elegantly in the STL with its set of iterator-based standard algorithms (\stdsort, \stdfill, etc.).

DASH generalizes core underlying STL concepts: Data containers can span the memory of multiple nodes and global iterators can refer to anywhere in this virtual global address space. It is thus natural to support parallel DASH equivalents for STL algorithms. An example for this is \lst{dash::min_element()} which is passed two global iterators to delineate the range for which the smallest element is to be found. While conceptually simple, a manual implementation of \lst{min_element} for distributed data can actually be quite tedious, let alone repetitive.

The DASH algorithm building blocks are collective, i.e., all units that hold data in the supplied range participate in the call. The algorithms work by first operating locally and then combining results as needed. For \lst{min_element} the local minimum is found using \lst{std::min_element()} and then the global minimum is determined using a collective communication operation and finally the result is broadcast to all participating units.
The beauty of the implementation lies in the fact that it will work in the same way with any arbitrary range in a DASH container, with any underlying data distribution and not just with simple data types but with composite data types as well.

The algorithms presently available in DASH include \lst{copy()}, \lst{copy_async()}, \lst{all_of()}, \lst{any_of()}, \lst{none_of()}, \lst{accumulate()}, \lst{transform()}, \lst{generate()}, \lst{fill()}, \lst{for_each()}, \lst{find()}, \lst{min_elment()}, and \lst{max_element()}.

\end{done} 

\section{Evaluation}\label{sec:Evaluation}

In this section we evaluate DASH on a number of benchmarks and perform a study with a
scientific C++ MPI proxy application. The platforms we use in our
study are as follows:

\paragraph{IBEX}
A two-socket shared memory system with Intel E5-2630Lv2 (Ivy
Bridge-EP) CPUs, 12 physical cores total, 15 MB L3 cache per CPU and
64 GB of main memory.

\paragraph{SuperMUC-HW}
Phase II of SuperMUC at the Leibniz Supercomputing Centre (LRZ)
consisting of 3072 nodes, each equipped with two E5-2697v3 CPUs
(Haswell-EP) with a total of 28 physical cores per node, 18 MB L3
cache per CPU and 64 GB of memory per node. The nodes are
interconnected by an Infiniband FDR14 network.

\subsection{Efficient Local Data Access}

\begin{figure}
\centering
\includegraphics[width=0.49\textwidth]{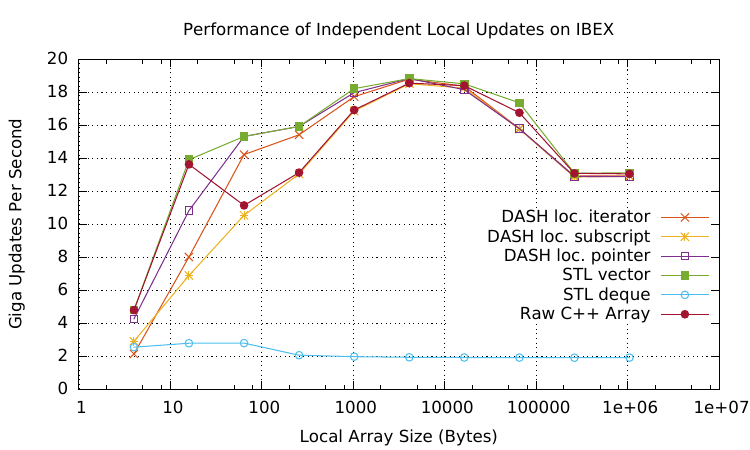}
\caption{Efficiency of local update operations in DASH using local subscripts, iterators and pointers, compared to raw array operations and the STL vector and deque.}\label{fig:IGUPS}
\end{figure}

This micro-benchmark tests how fast data can be accessed locally.  The
benchmark allocates a \dasharray of integers and initializes each
element to $0$. Then $N$ rounds of updates are performed, where in
each round each element is incremented by $1$. Each unit updates the
elements it owns (``owner computers'') and the total rate of updates
per second is reported by the benchmark in GUPS (giga updates per
second). Since data is only accessed locally, communication is not an
issue for this benchmark and we report data for a single node system
(IBEX).

Fig.~\ref{fig:IGUPS} shows the results achieved for this
benchmark. The horizontal axis shows different local array sizes while
the vertical axis plots the achieved update rate for a number of DASH
access variants and, for comparison, the \stdvector and \stddeque
container and a raw array (\lst{int[local\_size]}). For DASH we test
access to local data by local subscript (\lst{.local[i]}), local
iterator, and local pointer. As can be seen in Fig.~\ref{fig:IGUPS}, the
performance of all these access methods closely matches the
performance of the raw array accesses and the very well
performing \stdvector case. \stddeque shows much lower performance
because this container is not a optimized for access through the
subscript operator.

\subsection{Algorithms and Scalability}

\begin{figure}
\centering
\includegraphics[width=0.49\textwidth]{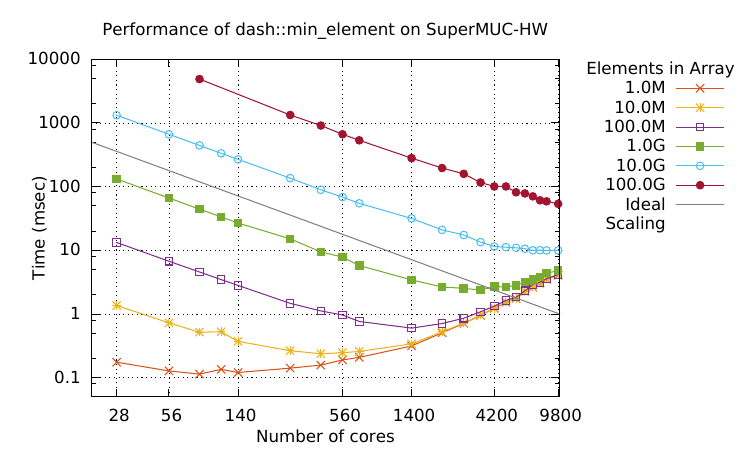}
\caption{Performance of the DASH min\_element() algorithm on arrays of up to 100 giga elements (400 GB) and up to 9800 cores.}\label{fig:min-element}
\end{figure}

In this benchmark we evaluate the performance and scalability of the
DASH algorithm \dashminelement{}.

The chart in Fig.~\ref{fig:min-element} shows the performance
of \dashminelement{} on SuperMUC. A \lst{dash::Array<int> arr} of
varying size is allocated using an increasingly large number of cores,
and \dashminelement{arr.begin(), arr.end()} is called. I.e., the
smallest element in the whole range is found.  The topmost line represents an
array of 100 billion entries (i.e., 400 GB of data) and the algorithm
scales nicely up to 9800 cores (350 nodes) of SuperMUC. At the largest
scale, finding the smallest entry takes about 50
milliseconds. For smaller arrays, the performance is dominated by communication and larger core counts increase the runtime.

\subsection{Communication Intensive Applications -- NPB DT}
The NAS Parallel Benchmarks DT kernel is a communication intensive benchmark
where a data flow has to be processed in a task graph. While the initial data
sets are randomly generated, the task graphs are quad trees with a binary
shuffle.  Depending on the problem size, the data sets fit into the L1-cache
(class S) and grow with higher problem classes. Since the task graph requires
frequent synchronization between the tasks, the crucial factor is a high
communication throughput. In the DASH implementation, the data sets are placed
in a globally distributed DASH Array which enables to use the
\lst{dash::copy_async} algorithm for the large data transfers. Due to the
efficient one-sided put operations we achieve up to 24\% better performance on
the SuperMUC, compared to the native MPI implementation:
\begin{table}[h]
\begin{tabular}[c]{ccrrrr}
\hline
Class & Graph & Size & Mop/s MPI   &  Mop/s DASH & Speedup\\
\hline
\hline
A & BH & 442368  & 170.80 & 175.16 & 1.03\\
A & SH & 442368  & 430.50 & 486.33 & 1.13\\
A & WH & 442368  & 313.02 & 387.47 & 1.24\\
B & BH & 3538944 & 210.34 & 215.02 & 1.02\\
B & SH & 3538944 & 776.38 & 905.96 & 1.17\\
B & WH & 3538944 & 463.20 & 459.91 & 0.99\\
\hline
\end{tabular}
\end{table}


\subsection{LULESH}

\begin{figure}
\centering
\includegraphics[width=0.49\textwidth]{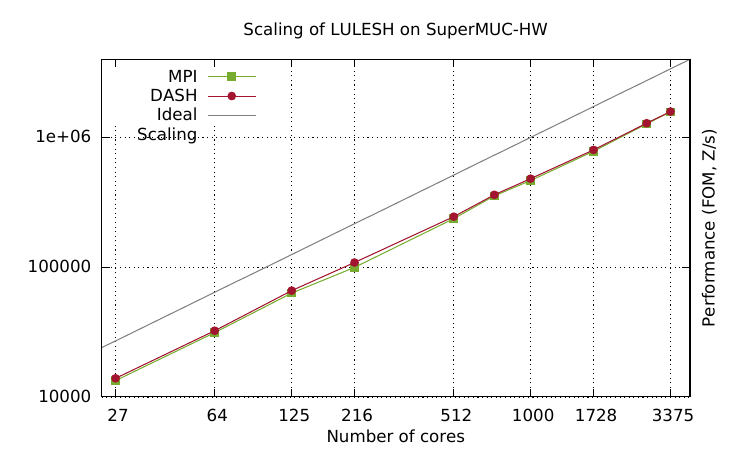}
\caption{Performance and scalability (weak scaling) of LULESH, implemented in MPI and DASH.}\label{fig:LULESH}
\end{figure}

Livermore Unstructured Lagrangian Explicit Shock Hydro-dynamics (LULESH) is a
widely used proxy application for calculating the Sedov blast
problem~\cite{Karlin:2013:LULESH} that highlights the performance
characteristics of unstructured mesh applications.

We ported LULESH v2.0.3 to DASH and perform a comparative performance and
scalability study. The Livermore MPI implementation keeps data (coordinates,
velocities, accelerations, etc.) in \stdvector containers and uses two-sided
message passing to exchange data along connecting faces, edges, and corners
with up to 26 neighbors in a 3D cube domain decomposition.

For our DASH port we place all data in globally distributed 3D \dashmatrix
containers, arrange units in a 3D cubic topology and use data distribution
scheme that is \blocked in each dimension. For a cubic number of processes this
results in the same decomposition used in the original version but requires far
less application-side code (index calculations, etc.), since DASH takes care of
these details. DASH also has the advantage that the data
distribution is not limited to cubic number of processes ($n^3$) but any number
$n\times m\times p$ of units can be used. We further replaced all two-sided
communication operations in the original version with asynchronous one-sided
put operations (using \lst{dash::copy_async}) that directly update the target
unit's memory. Fig.~\ref{fig:LULESH} shows the performance and scalability
comparison (using weak scaling) of the two versions on up to 3375 cores. DASH
scales similarly well as the optimized MPI implementation and offers
performance advantages of up to 9\%.


\section{Related Work}\label{sec:RelatedWork}

\begin{done}
The majority of scalable HPC applications use the message passing programming model in the form of MPI today. However, several factors will make it problematic to scale MPI to future extreme scale machines. First, MPI applications usually require some degree of data replication, which conflicts with the trend of continually shrinking available memory per compute core~\cite{Kogge:2013:Exascale}. Second, MPI applications typically implement a bulk-synchronous execution model and a more locally synchronized dynamic task-based execution approach is not easily realized in this model. Third, while MPI forces the programmer to think about data locality and thus leads to well performing code, it doesn't make it easy to work with data in a productive way -- e.g., by providing higher level data structure abstractions and supporting a global view across compute nodes.
\end{done}

\begin{done}
PGAS approaches make working with data in the above sense easier. PGAS essentially brings the advantages of threading-based programming (such as global visibility and accessibility of data elements) to distributed memory systems and accounts for the performance characteristics of data accesses by making the locality properties explicitly available to the programmer. Traditional PGAS approaches come in the form of a library (e.g., OpenSHMEM~\cite{Chapman:2010:Introducing}, Global Arrays~\cite{Nieplocha:1994:GA}) or language extensions (Unified parallel C, UPC~\cite{UPC:2005:Spec}, Co-Array Fortran, CAF~\cite{Mellor-Crummey:2009:CAF2, Numrich:1998:CAF}). Those solutions usually don't address hierarchical locality and offer only a two-level (local/remote) distinction of access costs. In contrast, DASH offers the concept of teams that can be used to express hierarchical organization of machines or algorithms.

Most programming systems also offer only \onedim arrays as their basic data-type out of which more complex data structures can be constructed -- but that work falls on the individual programmer. More modern PGAS languages such as Chapel~\cite{Chamberlain:2007:PPC} and X10~\cite{Charles:2005:X10} address hierarchical locality (e.g., in the form of locales or hierarchical place trees~\cite{Yan:2009:HPT}) but using these approaches requires a complete re-write of the application. Given the enormous amounts of legacy software, complete rewrites of large software packages are unlikely to happen. In contrast, DASH offers an incremental path to adoption, where individual data structures can be ported to DASH while leaving the rest of the application unchanged.
\end{done}

\begin{done}
Data structure libraries place the emphasis on providing data containers and operations acting on them. Kokkos~\cite{Edwards:2014:Kokkos} is a C++ template library that realizes \mdim arrays with compile-time polymorphic layout. Kokkos is an efficiency-oriented approach trying to achieve performance portability across various manycore architectures. While Kokkos is limited to shared memory nodes and does not address multi-level machine organization, a somewhat similar approach is followed by Phalanx~\cite{Garland:2012:Phalanx}, which also provides the ability to work across a whole cluster using GASNet as the communication backend. Both approaches can target multiple back-ends for the execution of their kernels, such as OpenMP for the execution on shared memory hardware and CUDA for execution on GPU hardware. RAJA~\cite{Hornung:2014:RAJA} and Alpaka~\cite{Zenker:2016:Alpaka} similarly target multiple backends for performance portability on single shared-memory systems optionally equipped with accelerators. RAJA, Alpaka and Kokkos are all restricted to a single compute node while DASH focuses on data structures that span the memory of many compute nodes.

STAPL~\cite{Buss:2010:STAPL} is a C++ template library for distributed data structures supporting a shared view programming model. STAPL doesn't appear to offer a global address space abstraction and can thus not be considered a bona-fide PGAS approach but it provides distributed data structure and a task-based execution model. STAPL offers flexible data distribution mechanisms that do however require up to three communication operations involving a directory to identify the home node of a data item. PGAS approaches in HPC usually forgo the flexible directory-based locality lookup in favor of a statically computable location of data items in the global address space for performance reasons. STAPL appears to be a closed-source project not available for a general audience.

Recently, C++ has been used as a vehicle for realizing a PGAS approach in the UPC++~\cite{Zheng:2014:UPC++} and Co-array C++~\cite{Johnson:2013:CoarrayCPP} projects. Co-array C++ follows a strict local-view programming approach and is somewhat more restricted than DASH and UPC++ in the sense that it has no concept of teams to express local synchronization and communication. While our previous work on the DASH runtime is based on MPI, UPC++ is based on GASNet. DASH offers support for hierarchical locality using teams, which are not supported by UPC++ and DASH more closely follows established C++ conventions by providing global and local iterators. STL algorithms can thus be applied directly on DASH data containers, which is not possible in UPC++. DASH also comes with a set of optimized parallel algorithms akin to those found in the STL while UPC++ offers no such algorithms. DASH also supports globalview \mdim distributed arrays with flexible data distribution schemes, UPC++ only supports a local view \mdim array inspired by Titanium~\cite{Kamil:2014:LocalView}.

\end{done}

\section{Conclusion and Future Work}\label{sec:Conclusion}
We have presented DASH, a compiler-free PGAS approach implemented as a C++ template library. Using one-sided communication substrates such as MPI-3 RMA, DASH offers distributed memory data structures that follow established C++ STL conventions and thus offer an easy adoption path for C++ developers. DASH can be integrated into existing applications by porting individual data structures at a time. DASH simplifies working with multidimensional data by providing a \mdim array abstraction with flexible data distribution schemes. DASH also accounts for increasingly complex machine topologies by providing a Team concept that can be used to express hierarchical locality. Our experimental evaluation has shown that DASH scales well and is able to outperform classic two-sided MPI applications.

DASH is free open-source software, released under a BSD license and is under active development. The current version of the library can be downloaded from the project's webpage at \url{http://www.dash-project.org}.

Further developments for DASH are planned in several directions. First, work is under way to support dynamically growing and shrinking containers. While classical HPC applications can typically be well supported by the fixed-size containers currently implemented in DASH, data analytics applications and HPC use cases in less traditional areas such as computational biology often require these more complex data structures.
Second, DASH focuses on data-structures and provides efficient support for the owner-computes model but it doesn't currently offer a way to apply computation actions to data elements in a more general way. Work towards this functionality is planned by implementing a general task-based execution model in the runtime and the C++ template library.

\section*{Acknowledgements}

We would like to thank the members of the DASH team and we gratefully acknowledge funding by the German Research Foundation (DFG) through the German Priority Programme 1648 Software for Exascale Computing (SPPEXA).

\bibliographystyle{IEEEtran}
\bibliography{references}

\end{document}